\documentclass[11pt,a4paper]{llncs}
\usepackage[a4paper,hmargin=2cm,vmargin=2.8cm]{geometry}
\usepackage{times}
\usepackage{epsfig}
\usepackage{graphicx}
\usepackage{amssymb}
\usepackage{amsfonts}
\usepackage{amsmath}
\usepackage{booktabs}
\usepackage{pstricks}
\usepackage{multirow}
\usepackage{wrapfig}
\usepackage[magyar,english]{babel}
\usepackage[latin2]{inputenc}
\usepackage{t1enc}

\newcommand{\vecti}[3]{\ensuremath{#1_{#2,1},#1_{#2,2},\dots,#1_{#2,#3}}}
\newcommand{\m}[1]{\mbox{$#1$}}

\newcommand{\raisedtothreefourths}[1]{\ensuremath{{#1}^{3/4}}}

\newcommand{\Prob}[1]{\ensuremath{\Pr\left\{#1\right\}}}
\newcommand{\Expect}[1]{\ensuremath{{\mathbb E}\left\{#1\right\}}}
\newcommand{\sumin}{\ensuremath{\sum_{i=1}^n}}
\newcommand{\E}{\ensuremath{\mathcal{E}}}
\newcommand{\FqL}{\ensuremath{\mathbb{F}_q^L}}
\newcommand{\Fq}{\ensuremath{\mathbb{F}_q}}
\let\Oldproof\proof
\renewenvironment{proof}{\Oldproof}{\qed}

\newlength{\figurewidth}
\title{Secure Capacity Region for Erasure Broadcast Channels with Feedback}
\date{}
\author{László Czap \quad  Vinod M. Prabhakaran \quad Suhas Diggavi \quad  Christina Fragouli}
\institute{EPFL, TIFR, UCLA and EPFL}

\begin{document}
\maketitle

\begin{abstract}
  We formulate and study a cryptographic problem relevant to wireless:
  a sender, Alice, wants to transmit private messages to two
  receivers, Bob and Calvin, using unreliable wireless broadcast
  transmissions and short public feedback from Bob and Calvin. We ask,
  at what rates can we broadcast the private messages if we also
  provide (information-theoretic) unconditional security guarantees
  that Bob and Calvin do not learn each-other's message?  We
  characterize the largest transmission rates to the two receivers,
  for { any} protocol that provides unconditional security
  guarantees. We design a protocol that operates at any rate-pair
  within the above region, uses very simple interactions and
  operations, and is robust to misbehaving users.
\end{abstract}

\section{Introduction}
\label{sec:intro}

\begin{figure}
\center
\includegraphics[width=\figurewidth]{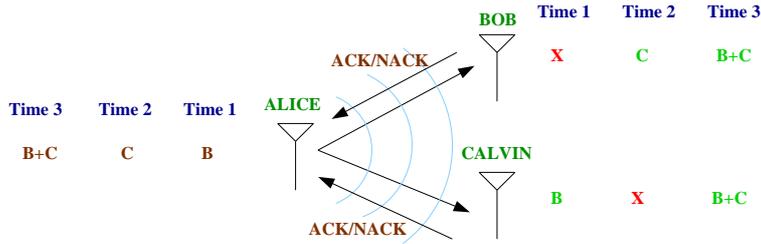}
\caption{An example where mixing private messages increases the
  transmission efficiency.  Alice wants to send packet B to Bob and C
  to Calvin; symbol X indicates that a transmitted packet is not
  successfully received due to corruption by the channel.  Alice first
  broadcasts message B; Calvin successfully receives it, while Bob
  fails to do so.  Next, Alice transmits message C; Bob successfully
  receives it, while Calvin fails. 
  Alice can take advantage of this side information Bob
  and Calvin have, and transmit the message B+C. This transmission is
  maximally useful for both receivers: assume both receive it, Bob,
  since he knows C, he can retrieve B, and Calvin, since he knows B,
  he can retrieve C. Thus the protocol concludes in three
  transmissions. In contrast, if we did not use coding across the
  private messages, we would need at least four transmissions, yielding  $33\%$ reduction
in  efficiency.}
\label{fig:MsgTx}
\vspace{-.5cm}
\end{figure}

Wireless bandwidth is scarce -- to efficiently use it, we need to mix
private messages intended for different users -- and this makes
securing such channels hard. Consider the situation where a wireless
access point, Alice, wants to send private messages to two receivers,
Bob and Calvin. To do so, Alice can only use the wireless channel,
where each packet transmission is broadcast and subject to errors.  A
simple strategy is for Alice to keep retransmitting each packet until
it is acknowledged by the intended receiver.  But as Alice repeatedly
broadcasts a packet intended for Bob, Calvin may overhear it. In fact,
recent work has established that Calvin {\em should} try to overhear
the packets intended for Bob, while Alice should code across the
private packets she has for Calvin and Bob, as this can significantly
increase the communication rates both receivers experience
\cite{Wu,Leandros09,MaddahAliTse10,Marios,Dana}.
\figurename~\ref{fig:MsgTx} illustrates such an example.
In the wireless community, the need for bandwidth efficiency is
acutely perceived, and there is significant effort in developing and
deploying such schemes that rely on opportunistic overhearing and
mixing of private messages
\cite{katti2008cope,more,Seferoglu_Markopoulou_2007}.  However, the
gain in efficiency seems to come with a security compromise, since Bob
and Calvin learn parts of each other's message. This leads us to ask a
new question. 

\paragraph{\bf Question:} In a wireless broadcasting
setting\footnote{To formalize this question we need to specify {(i)}
  what is a good model for the noisy broadcast channel?  {(ii)} what
  kind of feedback is feasible? {(iii)} what is the notion of security
  that we seek?  {(iv)} what is the power of the adversary? {(v)} what
  is the measure of transmission efficiency?  We formally discuss
  these aspects in Section~\ref{sec:SysModel}.}, can we characterize
the optimal unconditional (information-theoretic) secure transmission
rates for conveying private messages to Bob and Calvin?



In this paper we answer this question when Alice can use a broadcast
erasure channel, while Bob and Calvin can send (public, reliable, and
authenticated) packet acknowledgments. That is, each transmission of
Alice is either perfectly received or completely lost by Bob and
Calvin independently from each other, and they can causally
acknowledge this fact.  Channels that perfectly erase packets do not
exist in nature; even if a packet is corrupted by noise it would still
be possible to extract some information from it.  However, recent
experimental results on wireless testbeds show that one can create
almost perfect erasures through careful insertion of interference and
appropriate coding \cite{Arxiv11,HeshamElGamal10}. This mechanism also
enables erasure channels with known erasure
probabilities.
Furthermore,
the results for erasure channels can serve as building blocks for
noisy wireless channels \cite{DanaGowarikar06,ADT11,JMDF11}.

If we do not insist on information theoretic security guarantees,
there exist today methods to answer this question; our work as far as
we know is the first to examine whether it is possible to provide
unconditional guarantees and at what rates.

\paragraph{\bf Our contributions:} We propose a low-complexity
two-phase protocol, which first efficiently generates secure keys and
then judiciously uses them for encryption exploiting the wireless
broadcast properties.  We prove our protocol is optimal by showing
a matching impossibility result: we show that no other scheme
can achieve better secure private transmission rates to Bob and Calvin. To
the best of our knowledge, this is the first result on
information-theoretic security where the optimal strategy has a natural
need for a two-phase protocol for secure message transmission.
Our impossibility result also introduces new
information-theoretic techniques that utilize a balance between generated
and consumed keys, although the bounds are true for {\em any} valid
security protocol which does not constrain it to be a two-phase scheme.

Our protocol is based on the following ideas. First, Alice-Bob and
Alice-Calvin may create unconditionally secure pairwise secret keys $K_B$ and
$K_C$ respectively, using
a fundamental observation of Maurer \cite{Maurer1993}: different receivers
have different looks on the transmitted signals, and we can build on these
differences with the help of feedback to create secret keys
\cite{Arxiv11,HeshamElGamal10}. For example, if Alice transmits random
packets through independent erasure channels with erasure probability
$0.5$, there would be a good fraction of them (approximately $25\%$) that
only Bob receives, and we can transform this common randomness between
Alice and Bob to a key $K_B$ using privacy amplification\footnote{In fact
this can be done using linear combinations of the received packets, thereby
allowing for a complexity that is polynomial in the number of transmitted
packets \cite{Arxiv11,KanukurthiReyzin10}.}
\cite{Maurer1993,Arxiv11,HeshamElGamal10,KanukurthiReyzin10}. A novel
aspect of our protocol is that the secure keys for both Bob and Calvin are
generated {\em simultaneously} using the same sequence of transmissions by
Alice, thus optimally utilizing wireless broadcasting.

A naive approach is to generate the secret keys $K_B$, $K_C$ with the same
size as the respective private messages and use them as one-time pads.
This is too pessimistic in our case: {\em Calvin is only going to receive a
fraction of the packets intended for Bob and thus we only need to create an
amount of key that allows us to protect against this fraction.}  To build
on this observation, feedback is useful; knowing which packets Bob has
successfully received (or not), allows us to decide what to transmit next,
so that we preserve as much secrecy from Calvin as possible; and
symmetrically for Calvin.
In the second phase of our protocol, we combine these ideas with a network
coding strategy \cite{Yeung,primer,Wu} that makes
transmissions maximally useful to both Bob and Calvin.
\figurename~\ref{fig:all_three} shows the benefits of our approach (which
achieves the secret message capacity) compared to the naive scheme.

Our protocol does not rely on both users operating honestly: even if we
assume that Calvin  misbehaves, for example by sending fake acknowledgments
(see Section \ref{sec:Protocol}), we can still provide the same security
guarantees and operational rate to the correctly behaving Bob.


\paragraph*{\bf Related work:}
Secure transmission of messages using noisy channel properties was
pioneered by Wyner \cite{Wyner75}, who characterized the secret
message capacity of wiretap channels. This led to a long sequence of
research on information-theoretic security on various generalizations
of the wiretap channel \cite{CK78,LiangMonograph}. Notably, when the
eavesdropper and legitimate channel are statistically identical, then
the wiretap framework yields no security. The fact that feedback can
give security even in this case was first observed for secret key
agreement by Maurer \cite{Maurer1993} and further developed by
Ahlswede-Csiszár \cite{AC93} -- but secure key agreement is not the same
as secure transmission of {\em specific} messages.  The wiretap
channel with secure feedback and its variants for message security
have been studied in \cite{Hesham,Kim}; some conclusive results are
developed in special cases when there is a secure feedback
inaccessible to the eavesdropper.  Security of private message
broadcasting {\em without feedback} has been studied in
\cite{LiangShamai}, where some conclusive results have been
established.  As mentioned earlier, the use of feedback and broadcast
for private message transmission, {\em without} security requirements
has been studied in \cite{Leandros09,MaddahAliTse10}.  We believe that
ours are the first conclusive results that use insecure (and very
limited) feedback for information-theoretic security of multiple
private messages. We use linear code constructions that superficially
seem similar to those in secret sharing \cite{ShamirSecShare}, but our
problem is different since we obtain secrecy for all erasure
probabilities, by leveraging feedback\footnote{Secret sharing would
  require that the number of erasures of the adversary is smaller than that of
  the legitimate receiver.}.  Another problem that is
  different but bears some similarities, is that of broadcast
  encryption \cite{BroadcastEncrypt,BroadEnc93}, where a group of
    users receive a common secret message, and the issue is to deal
    with key management when users unsubscribe.

\paragraph*{\bf Outline:} The rest of the paper is organized as follows. 
Section~\ref{sec:SysModel} describes the communication and security model, 
Section~\ref{sec:MainRes} gives our main
result  and  a simple example,
 Section \ref{sec:Protocol} formally describes our protocol, 
Section \ref{sec:analysis} contains the security analysis, 
Section~\ref{sec:capacity} establishes optimality  by proving an impossibility 
theorem and Section~\ref{sec:Disc} concludes by discussing several possible extensions. Detailed proofs are provided in the Appendices.

\section{Problem formulation and system model}
\label{sec:SysModel}

We consider a three party communication setting with one sender
(Alice) and two receivers (Bob and Calvin).  The goal of Alice is to
securely send private messages $W_1$ and $W_2$ to Bob and Calvin, such
that the receivers may not learn each other's messages.  



Alice  employs a memoryless erasure broadcast
channel defined as follows. The inputs of the channel are length $L$
vectors over \Fq, which we call sometimes packets. The $i$th input is denoted by $X_i$. The $i$th
output of the channel seen by Bob is $Y_{1,i}$, while the output seen
by Calvin is $Y_{2,i}$. The broadcast channel consists of two
independent erasure channels towards Bob and Calvin. We denote
$\delta_1$ the erasure probability of Bob's channel and $\delta_2$
that of Calvin's channel. More precisely,
\begin{align}
&\Pr\{Y_{1,i},Y_{2,i}|X_i\}=\Pr\{Y_{1,i}|X_i\}\Pr\{Y_{2,i}|X_i\},\nonumber\\
&\Pr\{Y_{1,i}|X_i\}=\begin{cases}
1-\delta_1, &Y_{1,i}=X_i\\
\delta_1, &Y_i=\perp,
\end{cases}\nonumber \quad \mbox{ and } \quad
\Pr\{Y_{2,i}|X_i\}=\begin{cases}
1-\delta_2, &Y_{2,i}=X_i\\
\delta_2, &Y_{2,i}=\perp,
\end{cases}\nonumber
\end{align}
where $\perp$ is the symbol of an erasure.

\paragraph*{\textbf{Assumptions:}} We assume that the receivers send
public acknowledgments after each transmission stating whether or not
they received the transmission correctly. By \emph{public} we mean
that the acknowledgments are available not only for Alice but for the
other receiver as well.\footnote{In a practical setting, we do not
  need to actually have a public error-free channel to send the
  acknowledgments: since these are very short packets, we can utilize
  a sufficiently strong error correcting code and send them through
  the noisy wireless channels.}  We assume that some authentication
method prevents the receivers from forging each other's
acknowledgments.  Also, we assume that {\em both} Bob and Calvin only
know each other's acknowledgment causally, after they have revealed
their own (we justify this in Section~\ref{sec:Disc} when we discuss
Denial-of-Service attacks).

Let $S_i$ denote the state of the channel in the $i$th
transmission, $S_i\in\{B,C,BC,\emptyset\}$ corresponding to the
 receptions ``Bob only'', ``Calvin only'', ``Both'' and
``None'', respectively. Further, $S_i^*$ denotes the state based on the
acknowledgments sent by Bob and Calvin. If both users report honestly,
then \m{S_i=S_i^*}. We denote as $S^i$ the vector that collects all the states up to 
the $i$th, {\em i.e.,} $S^i=[S_1 \ldots S_i]$, and similarly for~$S^{i*}$.

Beside the communication capability as described above, all users can
securely generate private randomness. We denote by $\Theta_A, \Theta_B$ and
$\Theta_C$ the private random strings Alice, Bob, and Calvin, respectively
have access to. All parties  have perfect knowledge of the communication
model.

\subsection{Security and reliability requirements}

An $(n,\epsilon,N_1,N_2)$ scheme sends $N_1$ packets to Bob and $N_2$ to Calvin using $n$  transmissions from Alice with error probability smaller than $\epsilon$. Formally:
\begin{definition}
\label{def:1}
An $(n,\epsilon,N_1,N_2)$ {\em scheme} for the two user message transmission
problem consists of the following components: (a) message alphabets ${\mathcal
W}_1={\mathbb F}_q^{LN_1}$ and ${\mathcal W}_2={\mathbb F}_q^{LN_2}$, (b)
encoding maps $f_i(.)$, $i=1,2,\ldots,n$, and (c) decoding maps $\phi_1(.)$ and
$\phi_2(.)$, such that if the inputs to the channel are
\begin{align}
X_i = f_i(W_1,W_2,\Theta_A,S^{\ast i-1}),\quad i=1,2,\ldots, n,
\label{eq:def1_1}
\end{align}
where $W_1\in{\mathcal W}_1$ and $W_2\in{\mathcal W}_2$ are arbitrary
messages in their respective alphabets and
$\Theta_A$ is the private randomness Alice has access to, then,
provided the receivers acknowledge honestly, their estimates after decoding
$\hat{W}_1 = \phi_1(Y_1^n)$ and $\hat{W}_2 = \phi_2(Y_2^n)$ satisfy
\begin{align}
\Pr\{\hat{W}_1 \neq W_1\} &< \epsilon,\text{ and} \label{eq:def1_2} \\
\Pr\{\hat{W}_2 \neq W_2\} &< \epsilon. \label{eq:def1_3} 
\end{align}
\end{definition}

\begin{definition}
\label{def:2}
An $(n,\epsilon,N_1,N_2)$ scheme is said to be {\em secure
against honest-but-curious users} if in case both receivers are honest and the input messages $W_1$ and $W_2$ are independent random
variables distributed uniformly over their respective alphabets, in
addition to conditions \eqref{eq:def1_2}-\eqref{eq:def1_3} the
following two conditions also hold:
\begin{align}
I(W_1;Y_2^nS^n\Theta_C) &< \epsilon \label{eq:def2_1a} \\
I(W_2;Y_1^nS^n\Theta_B) &< \epsilon \label{eq:def2_1b}.
\end{align}
\end{definition}

\paragraph*{\textbf{Malicious user:}} We will say that a user is {\em
  malicious} if the user can ({\em a}) select the marginal
distribution of the other user's message arbitrarily; his own message
is assumed to be independent of the other user's message and uniformly
distributed over his alphabet and the malicious user does not have
access to his own message, and ({\em b}) produce dishonest
acknowledgments as a (potentially randomized) function of all the
information he has access to when producing each acknowledgment (this
includes all the packets and the pattern of erasures he received up to
and including the current packet he is acknowledging and the
acknowledgments sent by the other user over the public channel up to
the previous packet).
We allow at most one user to be malicious.

\begin{definition}
\label{def:3}
An $(n,\epsilon,N_1,N_2)$ scheme is said to be {\em secure against a
malicious user}, if in case one of the receivers is malicious (as defined above),
the scheme guarantees for the other (honest) receiver decodability and security
as in definitions \ref{def:1} and \ref{def:2}. That is, if Calvin
 is the malicious user, (\ref{eq:def1_2}) and (\ref{eq:def2_1a}) are satisfied for Bob,
while if Bob is the malicious user, (\ref{eq:def1_3}) and (\ref{eq:def2_1b}) are satisfied for Calvin.
\end{definition}
Clearly a scheme which is secure against a malicious user is also
secure against honest-but-curious users since the malicious user may
choose the uniform distribution for the other user's message and
choose to acknowledge truthfully.

\paragraph*{\textbf{Secret message capacity region:}} The
communication rate $R_i$ towards receiver $i$ expresses the number of
message $W_i$ bits successfully and securely delivered to receiver $i$
per channel use\footnote{Channel use refers to Alice using the channel
  once to send one packet.}.  We are interested in characterizing all
rate pairs $(R_1,R_2)$ that our channel can support.
\begin{definition}
\label{def:4}
The rate pair $(R_1,R_2)\in \mathbb{R}_+^2$ is said to be {\em achievable},
if for every $\epsilon,\epsilon'>0$ there are $N_1$ and $N_2$ and a large
enough $n$ such that there exists an $(n,\epsilon,
N_1,N_2)$ scheme that is secure against a malicious user and\footnote{All
logarithms in this paper are to the base 2 unless otherwise specified.}
\begin{align}
R_1-\epsilon'<\frac{1}{n}N_1L\log q, \quad R_2-\epsilon'<\frac{1}{n}N_2L\log
q . \label{eq:def4_1}
\end{align}
The {\em secret message capacity region} $\mathcal{R} \subset \mathbb{R}_+^2$ is defined
as the set of all achievable rate pairs.
\end{definition}

\section{Main result}
\label{sec:MainRes}

\begin{figure}[t]
\vspace{-.5cm}
\center
\includegraphics[width=\figurewidth]{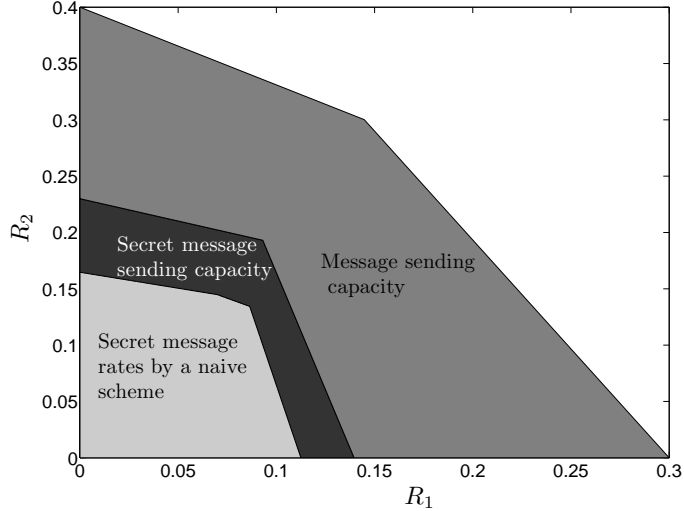}
\caption{Achieved rate region (in bits/packet) by the naive scheme in contrast to the optimal scheme and to the capacity region for the message sending problem with no security requirements. For this example \m{\delta_1=0.7}, \m{\delta_2=0.6}, \m{L=1}, \m{q=2}. Translating this region to bits/sec depends on how fast the source can send packets: for example, if we operate at the point ($R_1= 0.1$, $R_2=0.2$) and the  transmission rate of $1$Mbits/sec that IEEE802.11b supports, we would securely send $\approx100$Kbits/sec to Bob  and $\approx200$Kbits/sec to Calvin.}
\vspace{-.5cm}
\label{fig:all_three}
\end{figure}

%
%
\begin{theorem}
\label{thm:main}
The secret message capacity region as defined in Definition~\ref{def:4} is the set of all rate
pairs $(R_1,R_2)\in \mathbb{R}_+^2$ which satisfy the following two inequalities:
\begin{align}
\frac{R_1(1-\delta_2)}{\delta_2(1-\delta_1)(1-\delta_1\delta_2)}+\frac{R_1}{1-\delta_1}+\frac{R_2}{1-\delta_1\delta_2}&\leq
L\log q, \label{eq:th1_1}
\\ \frac{R_2(1-\delta_1)}{\delta_1(1-\delta_2)(1-\delta_1\delta_2)}+\frac{R_1}{1-\delta_1\delta_2}+\frac{R_2}{1-\delta_2}&\leq
L\log q. \label{eq:th1_2}
\end{align}
\end{theorem}
The first term of these inequalities can be interpreted as the
overhead for security, because -- as we will see soon -- it corresponds to the duration of a secret
key generation phase.  Omitting these terms gives us the capacity
region for the message transmission problem with two users without any
secrecy requirements~\cite{Leandros09}. The difference between these two capacity regions (with and without secrecy
requirements) is illustrated in \figurename~\ref{fig:all_three} for some specific values of the parameters $\delta_1$,
$\delta_2$, $L$ and $q$. 

We prove Theorem~\ref{thm:main} in two steps.  First, we provide a
protocol in Section~\ref{sec:Protocol} and prove in
Section~\ref{sec:analysis} that this protocol achieves all the rate
pairs in the capacity region. The complexity of the scheme is
discussed in Appendix~\ref{app:complexity}.  Then we provide in
Section~\ref{sec:capacity} a proof to show that (\ref{eq:th1_1}) and
(\ref{eq:th1_2}) are also bounds that are impossible to exceed by {\em
  any} protocol (a converse in information-theory
parlance). Interestingly, our converse holds even for the capacity
region defined using the weaker honest-but-curious security
definition, {\em i.e.,} a malicious user cannot deteriorate the
performance experienced by an honest user.  The following simple
example illustrates the main ideas in our protocol.
\subsection{A simplified example when both receivers are honest-but-curious}
\label{sec:Example}
Alice wants to securely send the $N_1=3$ message packets  $W_1=[W_{1,1},W_{1,2},W_{1,3}]$ to Bob and the $N_2=3$ message packets $W_{2}=[W_{2,1},W_{2,2},W_{3,2}]$ to Calvin.
Both Bob and Calvin are honest and report back truthfully. The protocol proceeds as depicted in Table~\ref{tab:example}.
\begin{table}
\vspace{-.4cm}
\caption{An example of a simplified protocol when both receivers are honest but curious.}
\vspace{-.5cm}
\label{tab:example}
\center
      \begin{tabular}{p{2.3cm}lcccccc}
	\toprule
	&\multirow{2}{*}{Alice sends}&			Bob's&		Calvin's&		\multirow{2}{*}{Bob's key}&	\multirow{2}{*}{Calvin's key}& \multirow{2}{*}{Bob decoded} &\multirow{2}{*}{Calvin decoded}\\
	&		&		ACK&		ACK\\
	\midrule
 	\multirow{5}{*}{\parbox{2cm}{Key generation} $\begin{cases}\\\\\\\\\end{cases}$}  & $X_1$ random&	$\checkmark$&	$\times$&      $K_{B,1}=X_1$&		\\
			& $X_2$ random&	$\checkmark$&	$\checkmark$&      $K_{B,1}$&		\\
			& $X_3$ random&	$\times$&	$\checkmark$&      $K_{B,1}$&	$K_{C,1}=X_3$&	\\
			& $X_4$ random&	$\times$&	$\checkmark$&      $K_{B,1}$&	$K_{C,1},K_{C,2}=X_4$&	\\
			& $X_5$ random&	$\checkmark$&	$\times$&      $K_{B,1},K_{B,2}=X_5$&	$K_{C,1},K_{C,2}$&	\\
 \dotfill&\dotfill&\dotfill&\dotfill&\dotfill&\dotfill &\dotfill&\dotfill   \\
	\multirow{3}{*}{\parbox{2cm}{Message transmission for Bob} $\begin{cases}\\\\\end{cases}$}&  $X_6=W_{1,1}\oplus K_{B,1}$& $\times$ &$\checkmark$ & $K_{B,2}$&$K_{C,1},K_{C,2}$& \\
			      &  $X_7=W_{1,2}\oplus K_{B,2}$& $\checkmark$ &$\times$ & $K_{B,2}$&$K_{C,1},K_{C,2}$& $W_{1,2}$\\
			      &  $X_8=W_{1,3}\oplus K_{B,2}$& $\checkmark$ &$\checkmark$ & &$K_{C,1},K_{C,2}$& $W_{1,2},W_{1,3}$\\
 \dotfill&\dotfill&\dotfill&\dotfill&\dotfill&\dotfill &\dotfill&\dotfill   \\
	\multirow{4}{*}{\parbox{2cm}{Message transmission for Calvin} $\begin{cases}\\\\\\\end{cases}$}&  $X_9=W_{2,1}\oplus K_{C,1}$& $\times$ &$\checkmark$ & &$K_{C,1},K_{C,2}$& $W_{1,2},W_{1,3}$&$W_{2,1}$\\ 
			      &  $X_{10}=W_{2,2}\oplus K_{C,1}$& $\times$ &$\times$ & &$K_{C,1},K_{C,2}$&  $W_{1,2},W_{1,3}$&$W_{2,1}$\\ 
			      &  $X_{11}=X_{10}$& $\checkmark$ &$\times$ & &$K_{C,2}$&  $W_{1,2},W_{1,3}$&$W_{2,1}$\\
			      &  $X_{12}=W_{2,3}\oplus K_{C,2}$& $\checkmark$ &$\times$ & &&  $W_{1,2},W_{1,3}$&$W_{2,1}$\\
     \dotfill&\dotfill&\dotfill&\dotfill&\dotfill&\dotfill &\dotfill&\dotfill   \\
	\multirow{2}{*}{\parbox{2cm}{Message sending for both} $\begin{cases}\\\end{cases}$}&  $X_{13}=X_6\oplus X_{11}$& $\checkmark$ &$\checkmark$ & &&  $W_{1,2},W_{1,3},W_{1,1}$&$W_{2,1},W_{2,2}$\\
				&  $X_{14}=X_{12}$& $\checkmark$ &$\checkmark$ & &&  $W_{1,2},W_{1,3},W_{1,1}$&$W_{2,1},W_{2,2},W_{2,3}$\\\hline 
      \end{tabular}      
\end{table}

\paragraph{Key generation:} Alice transmits five random packets
$X_1,\ldots,X_5$.  At the end of this phase, Alice and Bob share the
two secret key packets $K_{B,1}=X_1$ and $K_{B,2}=X_5$ that Bob
received and Calvin did not. Similarly, Alice and Calvin share the
secret key packets $K_{C,1}=X_3$ and $K_{C,2}=X_4$.  The packet $X_2$
which was received by both Bob and Calvin is discarded.

\paragraph{Message transmission for Bob:} 
We secure Bob's messages with one-time pads and transmit them until either Bob or Calvin receive them; we start by sending  $X_6=W_{1,1}\oplus K_{B,1}$, where $\oplus$ denotes addition in $\FqL$. \\
$-$ Since only Calvin receives $X_6$, we consider the key $K_{B,1}$ as {\em consumed} (Calvin observed a linear combination of it), and the  message $W_{1,1}$ as {\em undelivered} (Bob did not receive anything).  We will deal with undelivered messages at the last stage; 
at this point we proceed to send the new packet  $X_7=W_{1,2}\oplus K_{B,2}$.\\
$-$ Since only Bob receives $X_7$, we consider the message $W_{1,2}$ as delivered (Bob can retrieve it from $X_7$) and the key $K_{B,2}$ as unconsumed (it remains secret from Calvin); we can reuse it
 to send  $X_8=W_{1,3}\oplus K_{B,2}$.\\
$-$ Since  both Calvin and Bob receive $X_8$,  the message $W_{1,3}$ is delivered to Bob and the key $K_{B,2}$ is consumed.\\
At the end of this phase Bob has received packets $W_{1,2}$  and $W_{1,3}$ and is missing packet $W_{1,1}$.

\paragraph{Message transmission for Calvin:} We similarly make a first attempt to deliver Calvin's message.  We assume we are less successful than before, and although we consume both keys $K_{C,1}$ and $K_{C,2}$, we only deliver to Calvin $W_{2,1}$.  Note that $X_{10}$ which is not received by any user is simply retransmitted.

\paragraph{Message transmission for both:} To deliver the remaining messages $W_{1,1}$,  $W_{2,2}$ and $W_{2,3}$, we  take advantage of the fact that Bob already has $X_{11}$ and Calvin has $X_6$: we send $X_{13}=X_6\oplus X_{11}$ that is maximally useful for both. Bob can recover  $X_6$ and from this  $W_{1,1}$, while Calvin can recover $X_{11}$ and from this $W_{2,2}$. Note that $X_{13}$ brings no information to Bob or Calvin for each other's message.  

\paragraph*{\textbf{Important properties:}}
This simple scheme has the following properties.\\
$-$ The number of key packets we set up and consume is  smaller than the number of message packets we convey per user, because we can reuse certain keys that the adversary did not receive.\\
$-$ At the last message transmission phase, we exploit side information users have for each other's message to make a single transmission useful to both, without consuming  any new key.

\paragraph{Towards the general protocol:} If a node is dishonest he
can send fake acknowledgments. Interestingly, we can rely on the
expected behavior of the channel (and coding techniques) to have no
performance loss for the honest user.  For example, in the key
generation phase, if we expect that Calvin will only receive one of
the three packets Bob successfully receives, we can produce for Bob
the keys $K_{B,1}=X_1\oplus X_2$ and $K_{B,2}=X_2\oplus X_5$ that are
secure from Calvin no matter which packet he received.  Similarly,
when we send packets to Bob, assume we expect Calvin to receive two of
these transmissions -- but we do not know which two.  We then create
three linear combinations of Bob's keys, say $K'_{1,1}=K_{B,1}$,
$K'_{1,2}=K_{B,2}$, $K'_{1,3}=K_{B,1}\oplus K_{B,2}$, and transmit
$X_6=W_{1,1}\oplus K_{1,1}'$, $X_7=W_{1,2}\oplus K_{1,2}'$, and
$X_8=W_{1,3}\oplus K_{1,3}'$ - no matter which two of these Calvin
receives we are secure.  Our protocol builds on these ideas.

\section{Protocol}
\label{sec:Protocol}
We now describe a  $(n,\epsilon,N_1,N_2)$ scheme that is  secure against a malicious user as  in Definition~\ref{def:3}. 

\paragraph{\bf Parameters:} The operation of the protocol utilizes a set of parameters which we can directly calculate before the protocol starts, and whose use we will describe in the following.   
\begin{align}
k_B=N_1\frac{1-\delta_2}{1-\delta_1\delta_2} +
\raisedtothreefourths{\left(N_1\frac{1-\delta_2}{1-\delta_1\delta_2}\right)},&
\qquad
k_C=N_2\frac{1-\delta_1}{1-\delta_1\delta_2}+\raisedtothreefourths{\left(N_2\frac{1-\delta_1}{1-\delta_1\delta_2}\right)}.\label{eq:params-first}\\
 k_1=\frac{k_B}{\delta_2} + \frac{1}{\delta_2}
\left(\frac{2k_B}{\delta_2}\right)^{3/4},&\qquad
 k_2=\frac{k_C}{\delta_1} + \frac{1}{\delta_1}
\left(\frac{2k_C}{\delta_1}\right)^{3/4}.\\
n_1=\max\Bigg(\frac{k_1}{1-\delta_1}+\raisedtothreefourths{\left(\frac{k_1}{1-\delta_1}\right)},&\quad
              \frac{k_2}{1-\delta_2}+\raisedtothreefourths{\left(\frac{k_2}{1-\delta_2}\right)}\Bigg).\\
n_2 = \frac{N_1}{1-\delta_1\delta_2} +
\raisedtothreefourths{\left(\frac{N_1}{1-\delta_1\delta_2}\right)},&\qquad
n_3 = \frac{N_2}{1-\delta_1\delta_2} +
\raisedtothreefourths{\left(\frac{N_2}{1-\delta_1\delta_2}\right)}.\\
n_4 = \max\Bigg(\frac{N_1}{1-\delta_1}
+ \raisedtothreefourths{\left(\frac{N_1}{1-\delta_1}\right)} - n_2,& \quad \frac{N_2}{1-\delta_2}
+ \raisedtothreefourths{\left(\frac{N_2}{1-\delta_2}\right)} - n_3\Bigg).
\end{align}
\begin{align}
n=n_1+n_2+n_3+n_4.\label{eq:params-last}
\end{align}

\paragraph{\bf Main idea:} Using our protocol, Alice attempts to send $N_1$
message packets \m{W_1=(\vecti{W}{1}{N_1})} to Bob and
\m{W_2=(\vecti{W}{2}{N_2})} to Calvin using at most $n$ packet
transmissions. She either succeeds in sending $W_1$ to Bob or declares an
error for Bob. Similarly, she either succeeds in sending $W_2$ to Calvin or
declares an error for Calvin.  We will argue in Section \ref{sec:analysis}
that the failure probability can be made arbitrarily small.  She proceeds
in the following steps.
\begin{enumerate}
\item[I.] \emph{Key generation.} Generation of $k_B$ 
shared secret key packets between Alice-Bob and $k_C$ secret packets between Alice-Calvin using $n_1$ transmissions. This step fails if we do not succeed to generate the required number of secret key packets.
\item[II.] \emph{Message encryption and transmission.} She encrypts the messages using the produced keys and reliably transmits them to the two receivers. This step fails if we do not manage to deliver all the $N_1$ and $N_2$ message packets within the prescribed number of transmissions.
\end{enumerate}

\subsection*{Protocol Description}
{\em Key Generation}
\begin{enumerate}

\item Alice transmits $n_1$ packets $X_1,\ldots,X_{n_1}$. She  generates these packets uniformly at random from $\FqL$ using her private randomness, 
and independently of $W_1$, $W_2$.
\item Bob and Calvin acknowledge which packets they have received. If Bob receives less than $k_1$ packets we declare a protocol error for him. Similarly for Calvin if he receives less than $k_2$ packets. 
When an error is declared for both users, the protocol terminates. If not, we continue with the user not in error, as if the user in error did not exist.
\item  Let $X_1^B$  be a $L\times k_1$  matrix that has as columns the first $k_1$ packets that Bob
acknowledged.  Alice and Bob create $k_B$ secret key packets as
$K_B=X^B_{1}G_{K_B},$
where $G_{K_B}$ is a $(k_1\times k_B)$ matrix and is a parity check
matrix of an $(k_1,k_1-k_B)$ Maximum Distance Separable (MDS) code 
\cite{Sloane}. $K_B$ is a shared key set up
between Alice and Bob, and consists of $k_B$ length $L$ packets.

Similarly, using the first $k_2$ packets that that Calvin acknowledges,
Alice and Calvin create $k_C$ secret key packets.
The MDS codes are publicly known and fixed in advance. \\
\end{enumerate}
{\em Message encryption and transmission}\\
{\em Encryption}
\begin{enumerate}
\item[4.]  Alice and Bob produce $N_1$ linear combinations of their $k_B$ secret keys
as $K'_B=K_BG_{K'_{B}}$, where $G_{K'_{B}}$ is a
$(k_B\times N_1)$ matrix and is a generator matrix of an $(N_1,k_B)$ MDS
code which is also publicly known. Similarly, Alice and Calvin create $N_2$ 
linear combinations of their $k_C$ keys.
\item[5.] Alice creates $N_1$ encrypted messages to send to Bob  
\[
U_{B,i}=W_{1,i}\oplus K'_{B,i}, \quad i=1\ldots N_1
\]
where $\oplus$ is addition in the $\mathbb{F}_q^L$ vector space. Let $U_B$
denote the set of $U_{B,i},i=1,\ldots,N_1$. She similarly produces
a set $U_C$ of $N_2$ encrypted messages to send to Calvin
\[
U_{C,i}=W_{2,i}\oplus K'_{C,i}, \quad i=1\ldots N_2
\]
\end{enumerate}
{\em Transmissions to Bob}
\begin{enumerate}
\item[6.]  Alice sequentially takes each of the  $U_{B,i}$, $i=1\ldots N_1$, encrypted packets
and repeatedly transmits it, until it is acknowledged either by Bob or Calvin. That is,  if 
at time $i$ Alice transmits $X_i = U_{B,j}$ for some $j<N_1$, then 
\begin{align}
 X_{i+1}=\begin{cases}
X_{i}, & \text{if $S^\ast_{i}=\emptyset$}
\\ U_{B,j+1}, &\text{otherwise}.
\end{cases}
\end{align} 
\item[7.] (a) At the end of $n_2$ transmissions of $U_B$
packets, if Bob did not acknowledge $N_1(1-\delta_1)/(1-\delta_1\delta_2)$
of them, a protocol error is declared for Bob and we move to step 8 below.
If Calvin did not acknowledge $N_1(1-\delta_2)/(1-\delta_1\delta_2)$
packets, an error is declared for Calvin and continue transmitting $U_B$
packets.\\
(b) If all the $U_B$ packets are not exhausted before $n_2+n_4$
transmissions, we declare a protocol error for Bob and proceed to the next
step. 
\end{enumerate}
{\em Transmissions to Calvin}
\begin{enumerate}
\item[8.]  Similarly, Alice takes each of the  $U_{C,i}$, $i=1\ldots N_2$ packets
and repeatedly transmits it, until it is acknowledged either by Bob or Calvin.That is,  if 
at time $i$ we had $X_i = U_{C,j}$ for some $j<N_2$, then 
\begin{align}
 X_{i+1}=\begin{cases}
X_{i}, & \text{if $S^\ast_{i}=\emptyset$}
\\ U_{C,j+1}, &\text{otherwise}.
\end{cases}
\end{align}
\item[9.] (a) At the end of $n_3$ transmissions of $U_C$
packets, if Calvin did not acknowledge $N_2(1-\delta_2)/(1-\delta_1\delta_2)$
of them, a protocol error is declared for Calvin and we move to step 10 below.
If Bob did not acknowledge $N_2(1-\delta_1)/(1-\delta_1\delta_2)$
packets, we declare an error for Bob and continue transmitting $U_C$
packets.\\
(b) If all the $U_C$ packets are not exhausted before $n_3+n_4$
transmissions, we declare a protocol error for Calvin and proceed to the next
step. 
\end{enumerate}
{\em Transmissions to both}
\begin{enumerate}
\item[10.] At this step, Alice knows the following:
\begin{itemize}
\item 
At the end of step $6$, there may exist some encrypted packets that are
acknowledged {\em only by Calvin and not by Bob}.
Assume we have $N_1'$ such packets, and  denote them as $U_{B,i}'$, $i=1\ldots N_1'$. 
\item Similarly, 
at the end of step $8$, there may exist some $N_2'$ encrypted packets  $U_{C,i}'$, $i=1\ldots N_2'$,
that only Bob has acknowledged and not Calvin. 
\end{itemize} 
Alice proceeds to transmit $\oplus$ combinations of $U_{B,i}'$ and $U_{C,i}'$ packets.
\begin{itemize}
 \item She starts by transmitting  $U_{B,1}'\oplus U_{C,1}'$.
\item If at time $i$ Alice transmits $X_i = U_{B,j}'\oplus U_{C,\ell}'$ for some $j<N_1'$ and $\ell<N_2'$,
then
\begin{align}
X_{i+1}=\begin{cases}
  X_i  & \text{if $S_{i}^*=\emptyset$},\\
  U_{B,j+1}'\oplus U_{C,\ell}' & \text{if  $S_{i}^*=B$},\\
  X_{i+1}=U_{B,j}'\oplus U_{C,\ell+1}' & \text{if  $S_{i}^*=C$},\\
   X_{i+1}=U_{B,j+1}'\oplus U_{C,\ell+1}' &  \text{if  $S_{i}^*=BC$}.
\nonumber
\end{cases}
\end{align}
\item If Alice first finishes  all the $N_1'$ packets $U'_{B,i}$, she continues by transmitting the remaining  $N_2'$ packets until Calvin receives all of them.
\item Similarly if she first finishes all the $U'_{C,i}$ packets, she continues with the remaining $N_1'$ packets until Bob acknowledges them.
\end{itemize}
\item[11.] If steps 6 and 10 together exceed $n_2+n_4$ transmissions an error is
declared for Bob. Similarly, if steps 8 and 10 take together exceed
$n_3+n_4$ transmissions an error is declared for Calvin. In any case, the
protocol is terminated when the channel has been used for $n$ times.
\end{enumerate}

\section{Analysis}
\label{sec:analysis}
\begin{theorem}
For any $\epsilon,\epsilon'>0$ there exists a large enough $n$ for which
the scheme described above is secure against a malicious user and achieves
(in the sense of \eqref{eq:def4_1}) any rate pair in the region defined by
(\ref{eq:th1_1})-(\ref{eq:th1_2}).
\end{theorem}

\begin{proof}

Below, we prove that the above scheme is secure against a malicious user
and runs without error with high probability. The rate assertion of the
theorem follows from a simple numerical evaluation with the given parameter
values. 

\subsection{Security}

In our argument we focus on the secrecy of $W_1$ against a malicious
Calvin, but the same reasoning works for $W_2$ against a malicious Bob as
well.  Since we do not intend to give security guarantees to a malicious
user and consider at most one user to be malicious, we may assume that Bob
is honest. Moreover, under our definition of malicious user, $W_1$ and
$W_2$ are independent and the latter is uniformly distributed over its
alphabet, but the distribution of $W_1$ is arbitrary and controlled by the
malicious Calvin.

To analyze the secrecy of $W_1$, we may, without loss of generality, assume
that no error was declared for Bob during the key generation phase. Recall
that an error is declared for Bob only if Bob fails to acknowledge at least
$k_1$ packets. If an error was in fact declared for Bob, no information
about Bob's message $W_1$ is ever transmitted by Alice\footnote{More
precisely, if $E_\text{I-B}$ is the indicator random variable for an error
being declared for Bob in the key generation phase, \[ I(W_1;Y_2^n,S^n,\Theta_C)
\leq I(W_1;Y_2^n,S^n,\Theta_C,E_\text{I-B}) = I(W_1;Y_2^n,S^n,\Theta_C|E_\text{I-B})
\leq I(W_1;Y_2^n,S^n,\Theta_C|E_\text{I-B} = 0).\] To avoid clutter, we
leave out the conditioning event in the rest of this subsection.}. 
However, note that we do account for this error event when we analyze the
probability of error for Bob in the Section~\ref{sec:prob-of-error}.

We first show that $I(K_B;Y_2^{n_1}S^{n_1})$ can be made small, {\em
  i.e.,}~the key generation phase is secure.
\begin{lemma}
\label{lem:secret key generation}
When Bob is honest and no error is declared for Bob in the
key generation phase,
\begin{align}
I(K_B;Y_2^{n_1}S^{n_1})\leq k_Be^{-c_1\sqrt{k_1}}L\log q,\label{eq:secret
key leakage}
\end{align}
if $k_1= \frac{k_B}{\delta_2} +
\frac{1}{\delta_2}\raisedtothreefourths{\left(\frac{2k_B}{\delta_2}\right)}
$ and $k_B\geq \frac{2}{\delta_2}$,
where $c_1>0$ is some constant. Moreover, $K_B$ is uniformly distributed over its alphabet.
\end{lemma}
The key facts we use in proving this lemma are (i) the number of packets
seen by Calvin concentrates around its mean and (ii) an MDS parity check
matrix can be used to perform privacy amplification in the packet
erasure setting.

We still need to show that the secrecy condition (\ref{eq:def2_1a}) is
satisfied by the scheme even if Calvin controls the distribution of
$W_1$. We have
\begin{align}
I(W_1;Y_2^nS^n\Theta_C) \leq I(W_1;Y_2^nS^n\Theta_CU_C)
 = I(W_1;Y_2^n|Y_2^{n_1}S^n\Theta_CU_C), \label{eq:security-analysis1}
\end{align}
where the last equality used the fact that $\Theta_A,\Theta_C,W_2,S^n$ are
independent of $W_1$ and we may express $Y_2^{n_1},U_C$ as deterministic
functions of $\Theta_A,\Theta_C,W_2,S^n$.
Let ${1}_{B,i}^C$ be the indicator random variable for the
event that Calvin observes the packet $U_{B,i}$ either in its pure
form or in a form where the $U_{B,i}$ packet is added with some
$U_{C,j}$ packet. Let $M_B^C$ be the random variable which denotes the
number of distinct packets of $U_B$ that Calvin observes, so $M_B^C =
\sum_{i=1}^{N_1} 1_{B,i}^C.$ We have the following two lemmas:
\begin{lemma}
\label{lem:analysis:first-entropy-term}
$H(Y_2^n|Y_2^{n_1}S^n\Theta_CU_C) 
\leq \Expect{M_B^C} L\log q.$
\end{lemma}

\begin{lemma}
\label{lem:analysis:second-entropy-term}
$H(Y_2^n|W_1Y_2^{n_1}S^n\Theta_CU_C)\geq
\Expect{\min\left(k_B,M_B^C\right)} -
I(K_B;Y_2^{n_1}S^{n_1}).$
\end{lemma}
Using these in \eqref{eq:security-analysis1}, we have
\begin{align}
I(W_1;Y_2^nS^n\Theta_C) \leq \Expect{\max\left(0,M_B^C-k_B\right)} L\log q
+ I(K_B;Y_2^{n_1}S^{n_1}). \label{eq:security-analysis2}
\end{align}

Lemma~\ref{lem:secret key generation} gives a bound for the second
term. We can bound the first term using concentration inequalities. In
order to do this, let $Z_{B,i}$ be the number of repetitions of a
packet $U_{B,i}$ that Alice makes until Bob acknowledges it (where we
count both the transmission in pure form and in addition with some
packet from $U_C$). Note that the random variables $Z_{B,i}$ are
independent of each other and have the same distribution.  This
follows from the fact that the $S_i$ sequence is i.i.d., and each
$S_i$ is independent of $(Y_2^{i-1},S^{i-1},\Theta_C)$. In other
words, Calvin can exert no control over the channel state. Further,
for the same reason, with every repetition the chance that Calvin
obtains the transmission is $1-\delta_2$. This implies that the
indicator random variables $1_{B,i}^C$ are i.i.d. with
\[ \Prob{1_{B,i}^C=1} = (1-\delta_2) + \delta_1\delta_2(1-\delta_2) +
\ldots = \frac{1-\delta_2}{1-\delta_1\delta_2}.\]
Notice that $M_B^C$ is a sum of $N_1$ such independent random variables, and
hence
$\Expect{M_B^C} = N_1\frac{1-\delta_2}{1-\delta_1\delta_2}.$
Since
$k_B = N_1\frac{1-\delta_2}{1-\delta_1\delta_2} +
\raisedtothreefourths{\left(N_1\frac{1-\delta_2}{1-\delta_1\delta_2}\right)},
$
by applying Chernoff-Hoeffding bound we have
\begin{align}
\Expect{\max\left(0,M_B^C-k_B\right)} &\leq N_1 \Prob{M_B^C>k_B} \leq N_1
e^{-c_2\sqrt{N_1}}, \label{eq:ch_bound}
\end{align}
for a constant $c_2>0$. Substituting this together with
Lemma~\ref{lem:secret key generation} in \eqref{eq:security-analysis2} we
get
\begin{align*}
I(W_1;Y_2^nS^n\Theta_C) \leq N_1e^{-c_2\sqrt{N_1}} + k_Be^{-c_2\sqrt{k_B}},
\end{align*}
for constants $c_1,c_2>0$. By choosing\footnote{Recall from
\eqref{eq:params-first}-\eqref{eq:params-last} that by saying that we choose
$N_1$ large enough we cause $n$ to be large enough.} a large enough value
of $N_1$, we may meet \eqref{eq:def2_1a}.

\subsection{Error probability}\label{sec:prob-of-error}
We need to bound the probability that an error is declared for
Bob\footnote{Note that, under our protocol, if no error is declared for
Bob, he will be able to decode $W_1$.}. An error happens if:\\
$-$ Bob receives less than $k_1$ packets in the first phase,\\
$-$ he does not receive $N_1(1-\delta_1)/(1-\delta_1\delta_2)$ packets of
$U_B$ in step 7(a),
or $N(1-\delta_2)/(1-\delta_1\delta_2)$ packets of $U_C$ in step 9(a),\\
$-$ he does not receive all the $N_1$ packets of $U_B$ (either in pure
form or added with a packet in $U_C$) before step 11 intervenes.

All these error events have the same nature. An error happens if Bob
collects significantly fewer packets than he is expected to receive in a
particular step. The probability of these events can be bounded by applying
the Chernoff-Hoeffding bound as we did to show the security guarantee
(\ref{eq:ch_bound}). The sum of these bounds gives an upper bound on the
overall error probability of the scheme, which in turn can be made smaller
than $\epsilon$ by choosing $N_1$ large enough. 
A straightforward computation using the parameters in
\eqref{eq:params-first}-\eqref{eq:params-last} shows that \eqref{eq:def4_1}
is also satisfied.
\end{proof}
\section{Impossibility result (converse)}
\label{sec:capacity}
With Theorem~\ref{th3} we complete the proof of Theorem~\ref{thm:main}.
Throughout this section we will assume that both Bob and Calvin are honest.
Obviously, an upper bound for this case is a valid upper bound in the case
of a malicious user as well. Interestingly, we get the same bounds
for the honest-but-curious users' and for the malicious users' case. Our
proof relies on a few Lemmas which can be found together with their proofs
in Appendix~\ref{app:converse_lemmas}.

\begin{theorem}\label{th3} For the secret message capacity region as
  defined in Definition~\ref{def:4} it holds that:
\begin{align}
\frac{R_1(1-\delta_2)}{\delta_2(1-\delta_1)(1-\delta_1\delta_2)}+\frac{R_1}{1-\delta_1}+\frac{R_2}{1-\delta_1\delta_2}&\leq \nonumber
L\log q, 
\\ \frac{R_2(1-\delta_1)}{\delta_1(1-\delta_2)(1-\delta_1\delta_2)}+\frac{R_1}{1-\delta_1\delta_2}+\frac{R_2}{1-\delta_2}&\leq \nonumber
L\log q. 
\end{align}
\end{theorem}
\begin{proof}
  We will prove the first inequality, the second follows from
  symmetry. We look at Alice's transmissions from Bob's perspective
  and express them in three terms
  (\ref{eq:converse_terms}a)-(\ref{eq:converse_terms}c) using
  elementary properties of entropy:
\begin{align} &nL\log q\geq nH(X_i)\geq \sumin
H(X_i|Y_1^{i-1}S^{i-1})=\sumin\left [
H(X_i|Y_1^{i-1}Y_2^{i-1}S^{i-1})+I(X_i;Y_2^{i-1}|Y_1^{i-1}S^{i-1})\right
]\nonumber \\ &= \sumin\left [
\underbrace{H(X_i|Y_1^{i-1}Y_2^{i-1}S^{i-1}W_1)}_{(a)}
  +\underbrace{I(X_i;W_1|Y_1^{i-1}Y_2^{i-1}S^{i-1})}_{(b)}
+\underbrace{I(X_i;Y_2^{i-1}|Y_1^{i-1}S^{i-1})}_{(c)} \right ].
\label{eq:converse_terms} \end{align}
Lemmas~\ref{lemma:1}-\ref{lemma:6}
give lower bounds on each of the three terms in (\ref{eq:converse_terms});
putting these together results in the stated inequality.
\end{proof}

\paragraph{Intuition:} We now informally interpret the terms in
(\ref{eq:converse_terms}) with our protocol (of Section
\ref{sec:Protocol}) in mind. Note however that the proof provides a
general impossibility bound that holds for any scheme that satisfies
Definition~\ref{def:4}.  The terms
(\ref{eq:converse_terms}a)-(\ref{eq:converse_terms}c) classify the
information Alice sends during the $i$th transmission.  These terms
can also be interpreted as balancing key generation and consumption
for secrecy, as described below.

Term (\ref{eq:converse_terms}a) can be interpreted as anything that is
not related to Bob's message $W_1$ (and has not been seen by either
Bob or Calvin).  This is lower bounded through
Lemmas~\ref{lemma:5}-\ref{lemma:6}.  For example, in our protocol this
corresponds to a key generation attempt for either Bob or Calvin, or a
new encrypted message for Calvin, {\em i.e.,} steps 1 and 8. 

Term (\ref{eq:converse_terms}b) is interpreted as an encrypted packet
that Alice tries to send to Bob, which has not already been received
by Calvin.  This is lower bounded in Lemma~\ref{lemma:1}. In our
protocol this occurs in transmissions directed towards Bob only in
step 6.

Term (\ref{eq:converse_terms}c) brings information that Calvin has
already seen during previous transmissions, but Bob has not seen.
This is lower bounded in Lemma~\ref{lemma:4}. In our protocol this
would correspond to transmissions in step 10.

\section{Extensions and discussion} \label{sec:Disc}

We showed that it is possible to provide unconditional security guarantees
while wirelessly broadcasting two private messages; we characterized all
possible transmission rate pairs for the  private messages, and showed we can
achieve these using a simple protocol that efficiently generates and
utilizes an appropriate amount of key.  We conclude our paper with several
natural extensions and some open questions.

\par\noindent{\bf Practical deployment:} Our protocol has low complexity
(see Appendix A) and does not require changing the physical layer
transceivers of the three users; it is thus attractive for a potential
system deployment.  Although we only claim optimality under the modeling
assumptions of Section~\ref{sec:SysModel},  we believe  such a system could
enable operation at high secret message rates  (for the parameters in
Fig.~\ref{fig:all_three}, of the order of $100$Kbits/sec per user), using channel
conditioning techniques of \cite{Arxiv11,HeshamElGamal10}.

\par\noindent{\bf Common message:}
Assume that besides the private messages, we also have a common message $W_c$
(and corresponding rate $R_c$) that we want to deliver to both Bob
and Calvin. Our protocol and the converse proof can be easily
extended to cover this case. The capacity region  becomes 
\begin{multline}\label{eq:24}
  \max \left\{ \frac{R_1(1-\delta_2)}{\delta_2(1-\delta_1)(1-\delta_1\delta_2)},\frac{R_2(1-\delta_1)}{\delta_1(1-\delta_2)(1-\delta_1\delta_2)} \right\}\\
  +\max
  \left\{\frac{R_1+R_c}{1-\delta_1}+\frac{R_2}{1-\delta_1\delta_2},\frac{R_1}{1-\delta_1\delta_2}+\frac{R_2+R_c}{1-\delta_2}\right\}\leq
  L\log q,
\end{multline}
where the second term  is the known bound for (not-secure) message sending \cite{Leandros09}, while the first term corresponds to the
overhead of key generation, same as before.

\par\noindent{\bf Partially secret messages:} Another natural extension is
to keep secret only one  part of the private message to each user; that is,
we have $W_1=(W_1',W_1'')$, $W_2=(W_2',W_2'')$ with a secrecy requirement
only for $W_1'$ and $W_2'$. Accordingly, $R_1=R_1'+R_1''$,
$R_2=R_2'+R_2''$.  Assuming the messages are independent and $W_1'',W_2''$
are uniformly distributed over their alphabets, our results easily extends
to this case, with capacity region  
\begin{multline}
  \max \left\{ \frac{R_1'(1-\delta_2)}{\delta_2(1-\delta_1)(1-\delta_1\delta_2)}- \frac{R_1''}{1-\delta_1\delta_2},\frac{R_2'(1-\delta_1)}{\delta_1(1-\delta_2)(1-\delta_1\delta_2)} - \frac{R_2''}{1-\delta_1\delta_2},0 \right\}\\
  +\max
  \left\{\frac{R_1+R_c}{1-\delta_1}+\frac{R_2}{1-\delta_1\delta_2},\frac{R_1}{1-\delta_1\delta_2}+\frac{R_2+R_c}{1-\delta_2}\right\}\leq
  L\log q.
\end{multline}

\par\noindent{\bf Correlated erasures:} Our results extend to arbitrary
correlation between the erasure patterns, as long as the distribution is
known a-priori.  Both the protocol and the converse can be modified to
characterize the secure transmission rates for this case. The resulting
capacity region depends on the joint distribution.

\par\noindent{\bf Strengthening the malicious user:} Our security
guarantees assume that the malicious user may choose the marginal
distribution of the other user's message, but his own message is assumed to
be independent and uniformly distributed over its alphabet; moreover, he
can only learn his message  through the channel outputs  he receives. A
stronger  malicious user could choose the {\em joint distribution} of the
messages (and may also have access to his own message). Even under this
stronger security definition, it is not hard to see we can achieve nonzero
rates (e.g., by two instantiations of our protocol, first for Bob with
$R_2$ set to 0 and then for Calvin with $R_1$ set to 0), however we
conjecture that the capacity region is in general smaller than what we
derived here.

\par\noindent{\bf Denial-of-Service (DoS) attacks:} We leave the question
open if the malicious user launches denial-of-service attacks (outside
of what our current model allows); however, in general such attacks
can be deterred by ensuring they reveal who the attacker is. As an
example, in the key generation phase of our protocol, we assumed that
Bob \& Calvin cannot learn the other's feedback before sending their
own.  This assumption stops a malicious Bob from acknowledging the
exact same packets as Calvin which would lead to protocol failure for
Calvin and a DoS attack. In practice, for half the ACKs, we can ask
Bob to send them first before Calvin, and for the other half Calvin to
send them first before Bob, and thus identify users attempting such
attacks. In this category are also attacks that attempt to (partially)
control the channel, for example through physical layer jamming, where
we can resort to physical layer techniques to find the jammer's
real location.

\newpage
\bibliographystyle{splncs}
\bibliography{secrecy}
\appendix

\begin{table}
\caption{Summary of notation}
\label{tab:natations}
\center
      \begin{tabular}{cl}
	\toprule
	$X_i,Y_{1,i},Y_{2,i}$		&The $i$th input and outputs of the channel \\
	$S_i,S_i^*$			&The actual and the acknowledged $i$th state of the channel \\
	$\delta_1,\delta_2$		&Erasure probabilities of Bob's and Calvin's channel\\
	$W_1,W_2$			&Private messages for Bob and Calvin\\
	$K_B,K_C$			&Shared keys between Alice-Bob, Alice-Calvin \\
	$K_B',K_C'$			&Keys used for encryption, dependent linear combinations of packets in $K_B$ (or $K_C$) \\	
	$U_B,U_C$			&Encrypted messages of Bob and Calvin \\
	$L$				&Size of a packet in terms of \Fq\ symbols\\
	$N_1,N_2$			&Size of $W_1$ and $W_2$ (in packets)\\
	$R_1,R_2$			&Secret message rates for Bob and Calvin \\
	$\Theta_A,\Theta_B,\Theta_C$	&Private randomness of Alice, Bob and Calvin \\	
	$k_B,k_C$			&Size of the keys $K_B$ and $K_C$ (in packets)\\
	$V_i$				&$i$th element of any vector $V$\\
	$V^i$				&First $i$ elements of any vector
$V$, {\em i.e.,}~$(V_1,V_2,\ldots V_i)$\\
	\bottomrule
      \end{tabular}      
\end{table}

\section{{Complexity considerations}} \label{app:complexity}

It is clear from the analysis in Section~\ref{sec:analysis} that the
length $n$ of the scheme grows as
$\max\{O(\log^2(\frac{1}{\epsilon}),O(\frac{1}{{\epsilon'}^4})\}$, where
$\epsilon$ is the security and probability of error parameter, and 
$\epsilon'$ is the gap parameter associated with the rate
(see Definition~\ref{def:4}). The algorithmic complexity is quadratic
in $n$; quadratic from the matrix multiplication to produce the key.

Also, for the proposed scheme, the size (entropy in bits) of $\Theta_A$ is
linear in $n$ and no private randomness is needed at Bob and Calvin. For a
malicious user, we allow unlimited amount of private randomness. 

\section{Proof of lemmas in Section~\ref{sec:analysis}}

\setcounter{lemma}{0}
\begin{lemma}
When Bob is honest and no error is declared for Bob in the
key generation phase,
\begin{align}
I(K_B;Y_2^{n_1}S^{n_1})\leq k_Be^{-c_1\sqrt{k_1}}L\log q,
\end{align}
if $k_1= \frac{k_B}{\delta_2} + \frac{1}{\delta_2}
\raisedtothreefourths{\left(\frac{2k_B}{\delta_2}\right)}$ and $k_B\geq \frac{2}{\delta_2}$,
where $c_1$ is some constant. Moreover, $K_B$ is uniformly distributed over its alphabet.
\end{lemma}
\begin{proof}

With a slight abuse of notation, in the following $X_1^{BC}$ will denote
the \emph{actual} packets Calvin received (not necessarily the same as
those that he acknowledges) out of the first $k_1$ packets Bob received.
Note that here we assume that an error was not declared for Bob in the key
generation phase and hence Bob did receive at least $k_1$ packets in the
key generation phase.  Also let $X_1^{B\emptyset}$ be the packets seen only
by Bob among the first $k_1$ he receives.  Let $I_{B\emptyset}$ and
$I_{BC}$ be the index sets corresponding to $X_1^{B\emptyset}$ and
$X_1^{BC}$. Recall that $X_1^B$ denotes the first $k_1$ packets received by
Bob. The notation $M^{I}$ will denote a matrix $M$ restricted to the
columns defined by index set $I$. Given this,
\begin{align*}
I(K_B;Y_2^{n_1}S^{n_1})&=I(X_1^BG_{K_B};X_1^{BC}S^n)\\&=H(X_1^BG_{K_B})-H(X_1^BG_{K_B}|X_1^{BC}S^n)
\nonumber \\ &=k_BL\log q-H(X_1^BG_{K_B}|X_1^{BC}S^n) \\
&=k_BL\log q-H(\left[X_1^{B\emptyset}G_{K_B}^{I_{B\emptyset}}\quad
X_1^{BC}G_{K_B}^{I_{BC}}\right]|X_1^{BC}S^n)\\
&=k_BL\log q - H(X_1^{B\emptyset}G_{K_B}^{I_{B\emptyset}}|X_1^{BC}S^n)\\
&=k_BL\log q-H(X_1^{B\emptyset}G_{K_B}^{I_{B\emptyset}}|S^n),
\end{align*}
where the third equality follows from the MDS property of the matrix
$G_{K_B}$. Using the same property, we have
\begin{align*}
H(X_1^{B\emptyset}G_{K_B}^{I_{B\emptyset}}|S^n)&=\sum_{i=0}^{k_1}\min\{i,k_B\}L\log
q\Prob{|X_1^{B\emptyset}|=i} \\ 
&\geq k_BL\log q\sum_{i=k_B}^{k_1}\Prob{|X_1^{B\emptyset}|=i} = k_BL\log
q\Prob{|X_1^{B\emptyset}|\geq k_B} \\ 
&=k_BL\log q\left(1-\Prob{|X_1^{B\emptyset}|<k_B}\right) \\ 
&=k_BL\log q\left(1-\Prob{|X_1^{BC}|\geq k_1-k_B}\right) \\ 
&\stackrel{\text{(a)}}{\geq} k_BL\log q\left(1-\Prob{|X_1^{BC}|\geq
(1-\delta_2)k_1 + \raisedtothreefourths{k_1}}\right) \\ 
&\geq k_BL\log q
\left(1-\Prob{\left||X_1^{BC}|-E\left[|X_1^{BC}|\right]\right|>\raisedtothreefourths{k_1}}\right),
\end{align*}
where the inequality (a) follows from the fact that the conditions
on $k_B$ and $k_1$ imply that 
\[k_1-k_B \geq (1-\delta_2)k_1 + \raisedtothreefourths{k_1}.\]
The Chernoff-Hoeffding bound gives that for some constant $c_1>0$
\begin{align*}
\Prob{\left||X_1^{BC}|-E\left[|X_1^{BC}|\right]\right|>\raisedtothreefourths{k_1}}
\leq e^{-c_1\sqrt{k_1}}.
\end{align*}
So, we have that 
\begin{align}
I(K_B;Y^{n_1}S^{n})&\leq k_BL\log qe^{-c_1\sqrt{k_1}}.
\end{align}

The final assertion of the lemma is a simple consequence of the MDS
property of the code and the fact that $X^{n_1}$ are i.i.d. uniform.
\end{proof}

\begin{lemma}
\[H(Y_2^n|Y_2^{n_1}S^n\Theta_CU_C)
\leq \Expect{M_B} L\log q.\]
\end{lemma}
\begin{proof}
Let $U_B^C$ be a vector of length $N_1$
such that the $i$-th element $U_{B,i}^C$ is $U_{B,i}$ if Calvin observes
this $U_{B,i}$ either in the pure form 
or added with some element of $U_C$, and $U^C_{B,i}=\perp$ otherwise.
Let $1_{B,i}^C$ is the indicator random variable for the event
$U_{B,i}^C\neq\perp$.
It is easy to see that the following are information equivalent
({\em i.e.,} we can express each side as a deterministic function of the other)
\[ (Y_2^n,S^n,\Theta_C,U_C) \equiv (U_B^C,Y_2^{n_1},S^n,\Theta_C,U_C).\]
Therefore,
\[ H(Y_2^nS^n\Theta_CU_C) = H(U_B^CY_2^{n_1}S^n\Theta_CU_C).\]
\begin{align*}
H(Y_2^n|Y_2^{n_1}S^n\Theta_CU_C) &= H(U_B^C|Y_2^{n_1}S^n\Theta_CU_C)\\
&= \sum_{i=1}^{N_1} H(U_{B,i}^C|U_B^{C\,i-1}Y_2^{n_1}S^n\Theta_CU_C)\\
&= \sum_{i=1}^{N_1}
H(U_{B,i}^C|1_{B,i}^CU_B^{C\,i-1}Y_2^{n_1}S^n\Theta_CU_C)\\
&\leq \sum_{i=1}^{N_1} H(U_{B,i}^C|1_{B,i}^C)\\
&\leq \sum_{i=1}^{N_1} (L\log q) \Prob{1_{B,i}^C=1}\\
&= \Expect{\sum_{i=1}^{N_1} 1_{B,i}^C} (L\log q).
\end{align*}
where the third equality follows from the fact that the indicator random variable
$1_{B,i}^C$ is a deterministic function of the conditioning
random variables.

\end{proof}

\begin{lemma}
\[ H(Y_2^n|W_1Y_2^{n_1}S^n\Theta_CU_C) \geq 
\Expect{\min\left(k_B,M_B^C\right)} L\log q-
I(K_B;Y_2^{n_1}S^{n_1}).\]
\end{lemma}
\begin{proof}
We adopt the notation for $U_B^C$ and $1_{B,i}^C$ introduced in the proof
of Lemma~\ref{lem:analysis:first-entropy-term}. In addition, let ${K'}_B^C$
be defined in a similar manner as $U_B^C$ such that ${K'}_{B,i}^C=\perp$ if
$U_{B,i}^C=\perp$ and ${K'}_{B,i}^C={K'}_{B,i}$ otherwise. Also, let
$1_B^{C}$ be the vector of indicator random variables $1_{B,i}^C$,
$j=1,\ldots,N_1$.

Proceeding as in the proof of 
Lemma~\ref{lem:analysis:first-entropy-term}, we have
\begin{align*}
H(Y_2^n|W_1Y_2^{n_1}S^n\Theta_CU_C) 
&= H(U_B^C|W_1Y_2^{n_1}S^n\Theta_CU_C)\\
&=
H({K'}_B^C|W_1Y_2^{n_1}S^n\Theta_CU_C)\\
&\geq H({K'}_B^C|1_B^CW_1Y_2^{n_1}S^n\Theta_CU_C)\\
&= H({K'}_B^C|1_B^C) - I({K'}_B^C;W_1Y_2^{n_1}S^n\Theta_CU_C|1_B^C)
\end{align*}
But, from the MDS property of $G_{K'_B}$, and the fact that $K_B$ is
uniformly distributed over its alphabet, we have
\begin{align*}
H({K'}_B^C|1_B^C) &= \sum_{i=1}^{N_1}
\min(i,k_B)\Prob{\sum_{j=1}1_{B,j}^C=i} L\log q\\
&= \Expect{\min\left(k_B,\sum_{i=1}^{N_1} 1_{B,i}^C\right)} L \log q.
\end{align*} 
Also,
\begin{align*}
I({K'}_B^C;W_1Y_2^{n_1}S^n\Theta_CU_C|1_B^C)
&\stackrel{\text{(a)}}{=} 
I({K'}_B^C;Y_2^{n_1}S^{n_1}|1_B^C)\\
&\leq I({K'}_B^C1_B^C;Y_2^{n_1}S^{n_1})\\
&\leq I({K}_B;Y_2^{n_1}S^{n_1}).
\end{align*} 
where (a) follows from the fact that the distribution of $W_2$ (uniform and
independent of $S^n,\Theta_A,\Theta_C$) implies that $U_C$ is independent
of $\Theta_A,S^n$ and using this we can argue that 
the following is Markov chain
\[ {K'}_{B}^C  - (1_B^C,Y_2^{n_1},S^{n_1}) - (W_1,\Theta_C,U_C).\] 
Substituting back we have the lemma.
\end{proof}

\section{Proof of Lemmas~\ref{lemma:1}-\ref{lemma:6}}
\label{app:converse_lemmas}
First we give a bound on (\ref{eq:converse_terms}b). The first lemma expresses that Alice has to send sufficient information of message $W_1$ such that the Bob and Calvin together (in fact Bob himself also) can reconstruct it despite of erasures. 
\begin{lemma}
\label{lemma:1}
From conditions (\ref{eq:def1_1})-(\ref{eq:def1_3}) it follows that
\begin{align}
\sumin I(X_i;W_1|Y_1^{i-1}Y_2^{i-1}S^{i-1}) \geq \frac{nR_1}{1-\delta_1\delta_2}-\E_1
\nonumber \end{align}
where
$\E_1= \frac{h_2(\epsilon')+\epsilon' L\log q}{1-\delta_1\delta_2}$.
\end{lemma}
\begin{proof}
\begin{align}
&nR_1-\E_1(1-\delta_1\delta_2) \leq I(Y_1^nY_2^nS^n;W_1)=\sumin I(Y_{1i}Y_{2i}S_i;W_1|Y_1^{i-1}Y_2^{i-1}S^{i-1}) \nonumber \\
&=\sumin I(Y_{1i}Y_{2i};W_1|Y_1^{i-1}Y_2^{i-1}S^{i-1}S_i) =\sumin I(Y_{1i}Y_{2i};W_1|Y_1^{i-1}Y_2^{i-1}S^{i-1},S_i\neq \emptyset)\Pr\{S_i\neq\emptyset\} \nonumber \\
&=\sumin I(X_i;W_1|Y_1^{i-1}Y_2^{i-1}S^{i-1})(1-\delta_1\delta_2)
\nonumber \end{align}
Here, the first inequality is Fano's inequality \cite{CT91} (Chapter 2). Besides, we exploited the independence property of $S_i$.
\end{proof}

\begin{lemma}
\label{lemma:4}
From conditions (\ref{eq:def1_1})-(\ref{eq:def1_3}) it follows that
\begin{align}
\sumin I(X_i;Y_2^{i-1}|Y_1^{i-1}S^{i-1}) \geq \frac{nR_1\delta_1(1-\delta_2)}{(1-\delta_1)(1-\delta_1\delta_2)}-\E_2,
\nonumber \end{align}
\end{lemma}
where
$
\E_2= \frac{h_2(\epsilon')+\epsilon' L\log q}{1-\delta_1}.
$
\begin{proof}
From Lemma~\ref{lemma:3}
\begin{align}
&\frac{nR_1}{1-\delta_1}-\E_2\leq\sumin
  I(X_i;W_1|Y_1^{i-1}S^{i-1})\nonumber \\ &=\sumin
  I(X_i;W_1|Y_1^{i-1}Y_2^{i-1}S^{i-1})
  +I(X_i;Y_2^{i-1}|Y_1^{i-1}S^{i-1})-I(X_i;Y_2^{i-1}|Y_1^{i-1}S^{i-1}W_1)
  \nonumber \\ &\leq \sumin I(X_i;W_1|Y_1^{i-1}Y_2^{i-1}S^{i-1})+
  I(X_i;Y_2^{i-1}|Y_1^{i-1}S^{i-1}) \nonumber \\ &\leq
  \frac{nR_1}{1-\delta_1\delta_2} + \sumin
  I(X_i;Y_2^{i-1}|Y_1^{i-1}S^{i-1}) \label{eq:lemma3_8}
 \end{align}
To get (\ref{eq:lemma3_8}) we used Lemma~\ref{lemma:2}.
\end{proof}

Lemma~\ref{lemma:5} can be interpreted as the connection between the generation and consumption of the randomness Bob knows but Calvin doesn't. 
\begin{lemma}
\label{lemma:5}
From conditions (\ref{eq:def1_1})-(\ref{eq:def1_3}) it follows that
\begin{align}
\sumin H(X_i|Y_1^{i-1}Y_2^{i-1}S^{i-1}W_1) \geq \frac{(1-\delta_2)}{(1-\delta_1)\delta_2}\sumin I(X_i;Y_1^{i-1}|Y_2^{i-1}S^{i-1}W_1) \nonumber
\end{align}
\end{lemma}
\begin{proof}
\begin{align}
0&\leq{}
H(Y_1^nS^n|Y_2^nS^nW_1)={}H(Y_1^{n-1}S^{n-1}|Y_2^nS^nW_1)+H(Y_{1n}S_n|Y_1^{n-1}Y_2^nS^nW_1)
\nonumber \nonumber \\ &={}H(Y_1^{n-1}S^{n-1}|Y_2^{n-1}S^{n-1}W_1)-I(Y_1^{n-1}S^{n-1};Y_{2n}S_n|Y_2^{n-1}S^{n-1}W_1)+ H(Y_{1n}|Y_1^{n-1}Y_2^nS^nW_1) \nonumber
\nonumber \\ &={}H(Y_1^{n-1}S^{n-1}|Y_2^{n-1}S^{n-1}W_1) -I(Y_1^{n-1}S^{n-1};Y_{2n}|Y_2^{n-1}S^{n-1}S_nW_1) + H(Y_{1n}|Y_1^{n-1}Y_2^nS^nW_1) \nonumber
\nonumber \\ &={}H(Y_1^{n-1}S^{n-1}|Y_2^{n-1}S^{n-1}W_1) -I(Y_1^{n-1}S^{n-1};Y_{2n}|Y_2^{n-1}S^{n-1}W_1,C\subset
S_n)\Pr\{C\subset S_n\}\nonumber\nonumber \\ &\quad+
H(Y_{1n}|Y_1^{n-1}Y_2^nS^{n-1}W_1,S_n=B)\Pr\{S_n=B\} \nonumber
\nonumber \\&\quad+
H(Y_{1n}|Y_1^{n-1}Y_2^nS^{n-1}W_1,S_n=BC)\Pr\{S_n=BC\} \nonumber
\nonumber \\ &={}H(Y_1^{n-1}S^{n-1}|Y_2^{n-1}S^{n-1}W_1) -I(Y_1^{n-1}S^{n-1};X_n|Y_2^{n-1}S^{n-1}W_1)(1-\delta_2)\nonumber\nonumber
\\ &\quad+ H(X_n|Y_1^{n-1}Y_2^{n-1}S^{n-1}W_1)(1-\delta_1)\delta_2
+H(X_n|Y_1^{n-1}Y_2^{n-1}X_nS^{n-1}W_1)(1-\delta_1)(1-\delta_2)
\nonumber \nonumber \\ &={}H(Y_1^{n-1}S^{n-1}|Y_2^{n-1}S^{n-1}W_1)-I(Y_1^{n-1}S^{n-1};X_n|Y_2^{n-1}S^{n-1}W_1)(1-\delta_2)\nonumber
\nonumber \\&\quad+
H(X_n|Y_1^{n-1}Y_2^{n-1}S^{n-1}W_1)(1-\delta_1)\delta_2 \nonumber
\nonumber \end{align} We do the same steps recursively to obtain the
statement of the lemma.
\end{proof}

\begin{lemma}
\label{lemma:6}
From the conditions (\ref{eq:def1_1})-(\ref{eq:def1_3}) it also follows that
\begin{align}
\sumin I(X_i;Y_1^{i-1}|Y_2^{i-1}S^{i-1}W_1)+\E_5 >\frac{nR_1}{1-\delta_1\delta_2} + \frac{nR_2\delta_2(1-\delta_1)}{(1-\delta_2)(1-\delta_1\delta_2)} \nonumber
\end{align}
\end{lemma}
\begin{proof}
From Lemma~\ref{lemma:7},
\begin{align}
&\E_3 > \sumin I(X_i;W_1|Y_2^{i-1}S^{i-1}) \nonumber \\ &=\sumin
  H(X_i|Y_2^{i-1}S^{i-1})-H(X_i|Y_2^{i-1}S^{i-1}W_1) \nonumber
  \\ &=\sumin H(X_i|Y_1^{i-1}Y_2^{i-1}S^{i-1}) +
  I(X_1;Y_1^{i-1}|Y_2^{i-1}S^{i-1}) -H(X_i|Y_2^{i-1}S^{i-1}W_1)
  \nonumber \\ &=\sumin H(X_i|Y_1^{i-1}Y_2^{i-1}S^{i-1}W_1)
  -H(X_i|Y_2^{i-1}S^{i-1}W_1) \nonumber
  \\&\quad\quad+I(X_i;W_1|Y_1^{i-1}Y_2^{i-1}S^{i-1}) +
  I(X_1;Y_1^{i-1}|Y_2^{i-1}S^{i-1}) \nonumber \\ &=\sumin
  -I(X_i;Y_1^{i-1}|Y_2^{i-1}S^{i-1}W_1)+I(X_i;W_1|Y_1^{i-1}Y_2^{i-1}S^{i-1})
  + I(X_1;Y_1^{i-1}|Y_2^{i-1}S^{i-1}) \nonumber 
\end{align} 
From Lemma~1,
\begin{align}
\sumin I(X_i;W_1|Y_1^{i-1}Y_2^{i-1}S^{i-1}) &\geq \frac{nR_1}{1-\delta_1\delta_2} -\E_1. \nonumber \\
\nonumber \end{align}
Further, a symmetric result to Lemma~\ref{lemma:4} shows:
\begin{align}
\sumin I(X_i;Y_1^{i-1}|Y_2^{i-1}S^{i-1}) \geq
\frac{nR_2\delta_2(1-\delta_1)}{(1-\delta_2)(1-\delta_1\delta_2)}-\E_4,
\nonumber \end{align} where $ \E_4 = \frac{h_2(\epsilon')+\epsilon' L
  \log q}{1-\delta_2}.  $ Applying these bounds results the statement
of the lemma, with $\E_5=\E_3+\E_1+\E_4$.
\end{proof}

\begin{lemma}
\label{lemma:2}
From conditions (\ref{eq:def1_1})-(\ref{eq:def1_3}) it follows that
\begin{align}
\frac{nR_1}{1-\delta_1\delta_2} &\geq\sumin I(X_{i};W_1|Y_1^{i-1}Y_2^{i-1}S^{i-1})
\nonumber \end{align}
\end{lemma}
\begin{proof}
\begin{align}
&nR_1 \geq H(W_1) \geq I(Y_1^nY_2^nS^n;W_1) 
=\sumin I(Y_{1i}Y_{2i}S_i;W_1|Y_1^{i-1}Y_2^{i-1}S^{i-1}) \nonumber \\
&=\sumin I(Y_{1i}Y_{2i};W_1|Y_1^{i-1}Y_2^{i-1}S^{i-1}S_i) =\sumin I(Y_{1i}Y_{2i};W_1|Y_1^{i-1}Y_2^{i-1}S^{i-1},S_i\neq \emptyset) \Pr\{S_i\neq \emptyset \} \nonumber \\
&=\sumin I(X_{i};W_1|Y_1^{i-1}Y_2^{i-1}S^{i-1}) (1-\delta_1\delta_2)
\nonumber \end{align}
We used the same properties as before.
\end{proof}

With the same type of argument that we used to prove Lemma~\ref{lemma:1}, we can show also the following:
\begin{lemma}
\label{lemma:3}
From conditions (\ref{eq:def1_1})-(\ref{eq:def1_3}) it follows that
\begin{align}
\sumin I(X_i;W_1|Y_1^{i-1}S^{i-1}) \geq \frac{nR_1}{1-\delta_1}-\E_2.
\nonumber \end{align}
\end{lemma}

\begin{lemma}
\label{lemma:7}
From the security condition (\ref{eq:def2_1a}) it follows that
\begin{align}
\E_3 &> \sumin I(X_i;W_1|Y_2^{i-1}S^{i-1}),
\nonumber \end{align}
where
$
\E_3 = \frac{\epsilon}{1-\delta_2}.
$
\end{lemma}
\begin{proof}
From  (\ref{eq:def2_1a}), we have that
\begin{align}
&\epsilon>{} I(Y_2^nS^n\Theta_C;W_1)\geq I(Y_2^nS^n;W_1)=\sumin I(W_1;Y_{2i}S_i|Y_2^{i-1}S^{i-1}) ={} \sumin I(Y_{2i};W_1|Y_2^{i-1}S^{i-1}S_i)  \nonumber \\
&={} \sumin I(Y_{2i};W_1|Y_2^{i-1}S^{i-1},C\subset S_i)\Pr\{C\subset S_i\}  \nonumber \\
&={}\sumin I(X_i;W_1|Y_2^{i-1}S^{i-1},C\subset S_i)(1-\delta_2)={}\sumin I(X_i;W_1|Y_2^{i-1}S^{i-1})(1-\delta_2)
\nonumber \end{align}
\end{proof}

\end{document}